\documentclass[fleqn,usenatbib]{mnras}

\usepackage{newtxtext,newtxmath}
\usepackage{threeparttable}
\usepackage{multirow}
\usepackage{booktabs}
\usepackage{color}
\usepackage{bm}
\usepackage{adjustbox}
\usepackage{tablefootnote}

\usepackage[T1]{fontenc}

\DeclareRobustCommand{\VAN}[3]{#2}
\let\VANthebibliography\thebibliography
\def\thebibliography{\DeclareRobustCommand{\VAN}[3]{##3}\VANthebibliography}

\usepackage{graphicx}	
\usepackage{amsmath}	
\usepackage{amssymb}	

\title[Diffractive lensing of nHz GWs]{Diffractive lensing of nano-Hertz gravitational waves emitted from supermassive binary black holes by intervening galaxies}

\author[Ma et al.]{
Hao Ma$^{1,2}$,
Youjun Lu$^{1,2}$\thanks{luyj@nao.cas.cn},
Zhiwei Chen$^{1,2}$,
Yunfeng Chen$^{2,1}$
\\
$^{1}$\,National Astronomical Observatories, Chinese Academy of
Sciences, 20A Datun Road, Beijing 100101, China\\
$^{2}$\,School of Astronomy and Space Sciences, University of Chinese
Academy of Sciences, 19A Yuquan Road, Beijing 100049, China\\
}

\date{Accepted XXX. Received YYY; in original form ZZZ}

\pubyear{2023}

\begin{document}
\label{firstpage}
\pagerange{\pageref{firstpage}--\pageref{lastpage}}
\maketitle

\begin{abstract}
Pulsar timing array (PTA) experiments are expected to detect nano-Hertz gravitational waves (GWs) emitted from individual inspiralling supermassive binary black holes (SMBBHs). The GW signals from a small fraction of these SMBBHs may be diffractively lensed by intervening galaxies. In this paper, we investigate the diffractive lensing effects on the continuous GW signals from the lensed SMBBHs and estimate the detectable number of such signals by PTAs, such as the Chinese PTA (CPTA) and the Square Kilometer Array (SKA) PTA. We find that the amplitude of the lensed GW signals may be only amplified by a factor of $\sim 1.01-1.14$ ($16\%-84\%$ range) and the phase of the signals may shift somewhat due to the lensing, significantly different from those strongly lensed high frequency GW signals from compact binary mergers in the geometric optics. We estimate that $\sim 0.01\%$ of all detected nano-Hertz GW signals from individual SMBBHs by future PTA experiments are lensed by foreground galaxies (i.e., up to $\sim 106$ for CPTA and up to $\sim 289$ for SKA-PTA). However, the lensed nano-Hertz GW signals are difficult to be distinguished from those without lensing by the PTA observations only. We further discuss the possibility about the identification of the lensed nano-Hertz GW signals from SMBBHs via the electromagnetic detection of their host galaxies or active galactic nuclei. 
\end{abstract}




\section{Introduction}
\label{sec:intro}

Supermassive binary black holes (SMBBHs) can form in galactic centers as a natural consequence of frequent galaxy mergers \citep[e.g.,][]{BBR80nat, Yu=2002}. At the late stages of their inspiral (e.g., with period around a year or so), vast amount of gravitational wave (GW) radiations are emitted at the nano-Hertz (nHz) frequency band ($10^{-9} \sim 10^{-7} \, \rm{Hz}$), making them one of the main targets of the Pulsar Timing Arrays (PTAs). The ongoing International PTA \citep[IPTA,][]{IPTA} composed of the Parkes PTA \citep[PPTA,][]{PPTA}, the European PTA \citep[EPTA,][]{EPTA}, and the North-American Nanohertz Observatory for GWs \citep[NANOGrav,][]{NANOGrav}, are regularly updating their timing data set by monitoring a set of millisecond pulsars (MSPs) \citep{Arzoumanian=2018,Perera=2019,Kerr=2020,Antoniadis=2022}. Recently, new PTA experiments, including the Chinese PTA \citep[CPTA,][]{Lee=2016} and the Indian PTA \citep[InPTA,][]{Joshi=2018} are also joining the campaign for the detection of GW signals and MeerKAT PTA has released its first 2.5\,yr data \citep{Miles=2023}. Furthermore, the next-generation Very Large Array \citep[ngVLA,][]{Murphy=2018,Selina=2018} and the future PTA based on the Square Kilometer Array \citep[SKA,][]{Keane=2015,Janssen=2015} telescope will greatly improve the sensitivity of PTA observations by monitoring hundreds to a thousand of MSPs with very high timing precision.

The nHz GW signals from SMBBH inspirals can also be strongly gravitationally lensed by foreground galaxies like any other GW sources, including mergers of stellar binary black holes and binary neutron stars \citep[e.g.,][]{Biesiada=2014, Oguri=2018, LiSS=2018, MaHao=2023}. Since the nHz GW signals have long wavelength which are comparable with the Schwarzschild radius of the lens galaxies, the wave effect may be important for the gravitational lensing of these GW signals. As pointed out by \citet{Takahashi=2003}, the geometrical optics approximation breaks down and the wave effect must be taken into account if the lens mass $M_{\rm L}$ $\lesssim 10^{8} M_{\odot}/(f_{\rm GW}/\rm{mHz})$. For typical PTA sources ($10^{-9} \sim 10^{-7}\,\rm{Hz}$) lensed by foreground galaxies, the wave effect plays a significant role in the lensing when the lens mass $M_{\rm L} \lesssim (10^{12} \sim 10^{14}) M_{\odot}$. In this case, a lensed event only has one image rather than multiple images under the geometrical optics approximation, and the path of the lensed GW signal from the source to earth is different from those paths of the lensed multiple electromagnetic (EM) images from the same event \citep{Takahashi=2017,Morita=2019,Suyama=2020}. Understanding this multi-messenger propagation path (and also time delay) difference may be helpful in reconstructing the mass distribution of the lens galaxy \citep{Ezquiaga=2020}. In addition, the first detection of the lensed nHz GWs along with their EM counterparts may at the same time verify Einstein's general relativity again.

In this paper, we investigate the lensed GW signals from inspiralling cosmic SMBBHs in the wave optics regime as well as their lensed host galaxies based on the singular isothermal sphere (SIS) lens model. We estimate the number of the SMBBHs that can emit GW signals detected by future PTA experiments with different configurations and the detectability of their associated lensed host galaxies by the future sky survey Nancy Grace Roman Space Telescope \citep[RST;][]{Dore=2019}. We note that \citet{Khusid=2022} recently studied the GW signals from SMBBH inspirals being strongly lensed by adopting the geometrical optics approximation but without consideration of the wave effect. They found that the nHz GW signals from about $9-26$ lensed SMBBHs may be detected along with high chances to resolve their EM counterparts.

This paper is organized as follows. In Section~\ref{sec:2}, we describe the mock SMBBH sample generated from a cosmic formation and evolution model of SMBBHS in galaxy merger remnants \citep[for details see][]{ChenYF=2020}. Then we briefly introduce the wave effect in gravitational lensing of the nHZ GW signals from inspiralling SMBBHs and investigate how it affects the GW waveforms emitted from lensed cosmic SMBBHs in Section~\ref{sec:3}. Estimations of the detectable number of lensed nHz GW signals are given in Section~\ref{sec:4}. The detectability of the corresponding lensed host galaxies and/or EM counterparts are discussed in Section~\ref{sec:5}. Finally conclusions and discussions are given in Section~\ref{sec:6}. Throughout the whole paper, we adopted the $\rm{\Lambda CDM}$ cosmology model with $(h, \Omega_{\rm m}, \Omega_{\Lambda}) = (0.7, 0.3, 0.7)$.

\section{GW events from SMBBH inspirals}
\label{sec:2}

\subsection{GW samples}
\label{sec:SMBBH_sample}

To generate a mock sample for cosmic SMBBHs, we adopt the formation and evolution model of cosmic SMBBHs from \citet{ChenYF=2020}, which was in turn based on the study of the dynamical evolution of SMBBHs in galaxy merger remnants by \citet{Yu=2002}. In the model, the cosmic distribution of SMBBHs is described by $\Phi_{\rm BBH}(M_{\rm BH}, q_{\rm BH}, f_{\rm GW}, z)$, which is defined so that $\Phi_{\rm BBH}(M_{\rm BH}, q_{\rm BH}, f_{\rm GW}, z) dM_{\rm BH} dq_{\rm BH} df_{\rm GW}$ represents the comoving number density of SMBBHs at redshift $z$  with total mass in the range $M_{\rm BH}\rightarrow M_{\rm BH}+dM_{\rm BH}$, mass ratio in the range $q_{\rm BH}\rightarrow q_{\rm BH}+dq_{\rm BH}$ and GW frequency in the range $f_{\rm GW}\rightarrow f_{\rm GW}+df_{\rm GW}$; and the distribution function can be obtained through
\begin{equation}
\begin{aligned}
\Phi_{\mathrm{BBH}} &\left(M_{\mathrm{BH}}, q_{\mathrm{BH}}, f_{\mathrm GW}, z\right) \\
=& \int_0^t d t^{\prime} \int d M_* \int d q_* n_*\left(M_*, z^{\prime}\right) \mathcal{R}_*\left(q_*, z^{\prime} \mid M_*\right) \\
& \times p_{\mathrm{BH}}\left(M_{\mathrm{BH}}, q_{\mathrm{BH}} \mid M_*, q_*, z^{\prime}\right) \\
& \times p_f(f_{\mathrm GW}, t-t^{\prime} \mid M_*, q_*, M_{\mathrm{BH}}, q_{\mathrm{BH}}, z^{\prime}) \\
& \times P_{\text {intact }}\left(z, z^{\prime} \mid M_*\right).
\end{aligned}
\label{eq:SMBBH_ND}
\end{equation}
In the above equation, $n_*(M_*,z')$ is the stellar mass function of the spheroidal components of the merger remnants (which is also denoted as `bulge' for simplicity) so that $n_*(M_*,z')dM_*$ represents the comoving number density of bulges at redshift $z'$ with mass in the range $M_*\rightarrow M_*+dM_*$; $\mathcal{R}_*(q_*,z'|M_*)$ is the merger rate per bulge so that $\mathcal{R}_*(q_*,z'|M_*)dq_*$ represents the averaged number of mergers with mass ratio in the range $q_*\rightarrow q_*+dq_*$ within time $t'\rightarrow t'+dt'$ for a descendant bulge with mass $M_*$, where $t'$ is the corresponding cosmic time at redshift $z'$. $p_{\rm BH}(M_{\rm BH},q_{\rm BH}|M_*,q_*,z')$ is a probability distribution defined so that $p_{\rm BH}(M_{\rm BH},q_{\rm BH}|M_*,q_*,z') dM_{\rm BH} dq_{\rm BH}$ represents the probability that a host merger at redshift $z'$ characterized by $(M_*,q_*)$ leads to a SMBBH merger characterized by $(M_{\rm BH}, q_{\rm BH})$ with $M_{\rm BH}$ representing the total mass of the SMBBHs, which is determined by the MBH--host galaxy scaling relation. $p_f(f_{\rm GW},\tau| M_*, q_*, M_{\rm BH}, q_{\rm BH}, z')$ is a probability distribution defined so that $p_f(f_{\rm GW},\tau|M_*,q_*, M_{\rm BH}, q_{\rm BH}, z')df_{\rm GW}$ represents the probability that a host merger characterized by parameters $(M_*,q_*,M_{\rm BH},q_{\rm BH})$ at redshift $z'$ leads to a SMBBH emitting GW at frequency range $f_{\rm GW}\rightarrow f_{\rm GW}+df_{\rm GW}$ at a time $\tau$ after the host merger\footnote{Note that $p_a(a, \tau | M_*, q_*, M_{\mathrm{BH}}, q_{\mathrm{BH}}, z')$ was used by \citet{ChenYF=2020} since they pursued the distribution of the semimajor axis $a$ instead of the GW frequency $f_{\rm GW}$ as we do here. The two quantities $a$ and $f_{\rm GW}$ can be converted to each other through $f_{\mathrm{GW}} = \pi (1+z) (G M_{\mathrm{BH}}/a^3)^{1/2}$ assuming circular orbits.}; and it encodes the different dynamical processes the SMBBHs undergo from galaxy mergers to their own mergers (detailed considerations can be found in \citealt{ChenYF=2020} and \citealt{ChenYF=2023}). $P_{\rm intact}(z,z'|M_*)$ represents the probability that the host merger remnant will not experience a major merger with another galaxy or be accreted by a bigger galaxy to become a satellite galaxy of it between redshifts $z$ and $z'$.

In Equation~\eqref{eq:SMBBH_ND}, the mass function and merger rate of the bulges can be derived from quantities of their host galaxies, i.e., 
\begin{equation}
\begin{aligned}
n_* &(M_*, z^{\prime}) \mathcal{R}_*(q_*, z^{\prime} \mid M_*) \\
=& \int d M_{\mathrm{gal}} \int d q_{\mathrm{gal}} n_{\mathrm{gal}}(M_{\mathrm{gal}}, z^{\prime}) \mathcal{R}_{\mathrm{gal}}(q_{\mathrm{gal}}, z^{\prime} \mid M_{\mathrm{gal}}) \\
\quad & \times p_*(M_*, q_* \mid M_{\mathrm{gal}}, q_{\mathrm{gal}}, z^{\prime})
\end{aligned}
\end{equation}
where $n_{\rm gal}(M_{\rm gal,z})$ and $\mathcal{R}_{\rm gal}(q_{\rm gal},z|M_{\rm gal})$ are the galaxy stellar mass function and the merger rate per galaxy, respectively; $p_*(M_*,q_*|M_{\rm gal},q_{\rm gal},z)$ is the probability distribution of the bulge total mass and mass ratio given the host galaxy total mass and mass ratio. All these galaxy quantities are defined in a similar way as the bulges.

We generate a mock sample of cosmic SMBBHs with source parameters $(z, M_{\rm BH}, q_{\rm BH}, f_{\rm GW})$ based on Equation~\eqref{eq:SMBBH_ND}. We restrict the sample SMBBHs to the redshift range $0.2 \leq z\leq3$, the total mass range $10^7\leq M_{\rm BH}/M_\odot\leq 10^{11}$, the mass ratio range $q_{\rm BH}\geq 0.01$ and the GW frequency range $10^{-9}\leq f_{\rm GW}/{\rm Hz}\leq 10^{-7}$. We will discuss the parameter distribution of these mock SMBBH systems and the lensed ones in section~\ref{sec:4}. Throughout this study, we ignore the orbit eccentricity of the SMBBHs for simplicity, since SMBBHs generally do not form with high eccentricities at the end of the dynamical friction stage \citep{Polnarev=1994} and the eccentricity decays rapidly due to GW radiation \citep{Peter=1963,Baker=2006}. Nevertheless, there are still debates on the eccentricity of SMBBHs. The scattering of ambient stars and interaction with circumbinary disks may also lead to some eccentric SMBBHs \citep{Quinlan=1996,Armitage=2005,Roedig=2011}.

\subsection{SNR}
\label{sec:SNR}

Once a GW signal is passing by the propagation path of the pulses from a millisecond pulsar (MSP) to earth, fluctuations of the pulse time of arrivals (ToAs) is induced and therefore the GW signals can be extracted from the pulsar timing residuals. Considering a single continuous GW source located at a direction $\hat{\Omega}$, the induced timing residual on the ToAs can be calculated by \citep[e.g.,][]{Thorne=1987,Jenet=2006,Zhu=2015,Yan=2020}
\begin{equation}
s(t, \hat{\Omega})=F^{+}(\hat{\Omega}) \Delta A_{+}(t)+F^{\times}(\hat{\Omega}) \Delta A_{\times}(t).
\label{eq:time_residuals}
\end{equation}
In the above equation, the antenna pattern functions $F^{{+}}(\hat{\Omega})$ and $F^{{\times}}(\hat{\Omega})$ are given by
\begin{equation}
\begin{aligned}
F^{+}(\hat{\Omega})=&\frac{1}{4(1-\cos \theta)}\left\{\left(1+\sin ^{2} \delta\right) \cos ^{2} \delta_{\mathrm{p}} \cos \left[2\left(\alpha-\alpha_{\mathrm{p}}\right)\right]\right. \\
&\left. -\sin 2 \delta \sin 2 \delta_{\mathrm{p}} \cos \left(\alpha-\alpha_{\mathrm{p}}\right)+\cos ^{2} \delta\left(2-3 \cos ^{2} \delta_{\mathrm{p}}\right)\right\},
\end{aligned}
\end{equation}
\begin{equation}
\begin{aligned}
F^{\times}(\hat{\Omega})=&\frac{1}{2(1-\cos \theta)}\left\{\cos \delta \sin 2 \delta_{\mathrm{p}} \sin \left(\alpha-\alpha_{\mathrm{p}}\right)\right. \\
&\left. -\sin \delta \cos^2\delta_{\mathrm{p}} \sin \left[2\left(\alpha-\alpha_{\mathrm{p}}\right)\right]\right\},
\end{aligned}
\end{equation}
with $(\alpha,\delta)$ and $(\alpha_{\rm p},\delta_{\rm p})$ denoting the sky locations of the SMBBH and pulsar, respectively.  The angle between the source and pulsar directions  with respect to the observer is denoted as $\theta$, and $\cos \theta=\cos \delta \cos \delta_{\mathrm{p}} \cos \left(\alpha-\alpha_{\mathrm{p}}\right)+\sin \delta \sin \delta_{\mathrm{p}}$.
The GW induced timing residual $s(t, \hat{\Omega})$ is contributed by both the Earth term $A_{\{+,\times\}}(t)$ and the pulsar term $A_{\{+,\times\}}(t_{\rm p})$ as $\Delta A_{\{+,\times\}}(t)=A_{\{+,\times\}}(t)-A_{\{+,\times\}}(t_{\rm p})$, where $t_{\rm p}$ is the time when the GW signals pass by the pulsar and can be obtained as $t_{\rm p} = t-D_{\rm p}(1-\cos \theta)/c$, with $D_{\rm p}$ representing the distance of the pulsar.
$A_{\{+,\times\}}(t)$ is defined as $A_{\{+,\times\}}(t)= h_{\{+,\times\}}(t) / 2 \pi f_{\rm GW}(t)$, and $h_{\{+,\times\}}(t)$ can be described as 
\begin{equation}
\label{eq:h_p}
\begin{aligned}
h_{+} =  h_{0} \big\{(1+\cos^2 \iota) &\cos 2\psi \sin \left[\phi (t)+\phi_0\right]  \\ &  + 2\cos \iota \sin 2\psi \cos \left[\phi(t)+\phi_0\right] \big\},  
\end{aligned}
\end{equation}
\begin{equation}
\label{eq:h_t}
\begin{aligned}
h_{\times} = h_{0} \big\{(1+\cos^2 \iota) & \sin 2\psi \sin \left[\phi (t)+\phi_0\right] \\ & - 2\cos \iota \cos 2\psi \cos \left[\phi(t)+\phi_0\right] \big\}, 
\end{aligned}
\end{equation}
with $\iota$ representing the inclination angle of the normal of the binary orbit with respect to the line of sight. For simplicity, we set the polarization angle of GW $\psi=0$ and the initial phase angle $\phi_0=0$ in the following calculations, which does not affect the main results. The GW amplitude $h_{0}$ here is determined by the redshifted chirp mass $\mathcal{M}_{\rm c}^{z} = (1+z)\mathcal{M}_{\rm c} = (1+z)M_{\rm BH} q_{\rm BH}^{3/5}/(1+q_{\rm BH})^{6/5}$, the GW frequency in the observer rest frame $f_{\rm GW}$, and the luminosity distance of GW source $D_{\rm L}$ as 
\begin{equation}
h_0  =  2 \frac{\left(G \mathcal{M}_{\mathrm{c}}^{z}\right)^{5 / 3}}{c^{4}} \frac{\left(\pi f_{\rm GW} \right)^{2 / 3}}{D_{\mathrm{L}}}.
\label{eq:h0}
\end{equation}
Once the initial GW observational frequency $f_{\rm GW,0}$ and chirp mass $\mathcal{M}_{\mathrm{c}}^{z}$ are given, the GW frequency $f_{\rm GW}(t)$ and phase $\phi(t)$ at time $t$ can be expressed as
\begin{equation}
\label{eq:f_GW_evolution}
f_{\rm GW}(t)=\left[f_{\rm GW,0}^{-8 / 3}-\frac{256}{5} \pi^{8 / 3}\left(\frac{G \mathcal{M}_{\mathrm{c}}^{z}}{c^{3}}\right)^{5 / 3} t\right]^{-3 / 8},
\end{equation}
\begin{equation}
\phi(t)=\frac{1}{16}\left(\frac{G \mathcal{M}_{\mathrm{c}}^{z}}{c^{3}}\right)^{-5 / 3}\left\{\left(\pi f_{\rm GW,0}\right)^{-5 / 3}-[\pi f_{\rm GW}(t)]^{-5 / 3}\right\}.
\end{equation}

\begin{figure}
\centering
\includegraphics[width=\columnwidth]{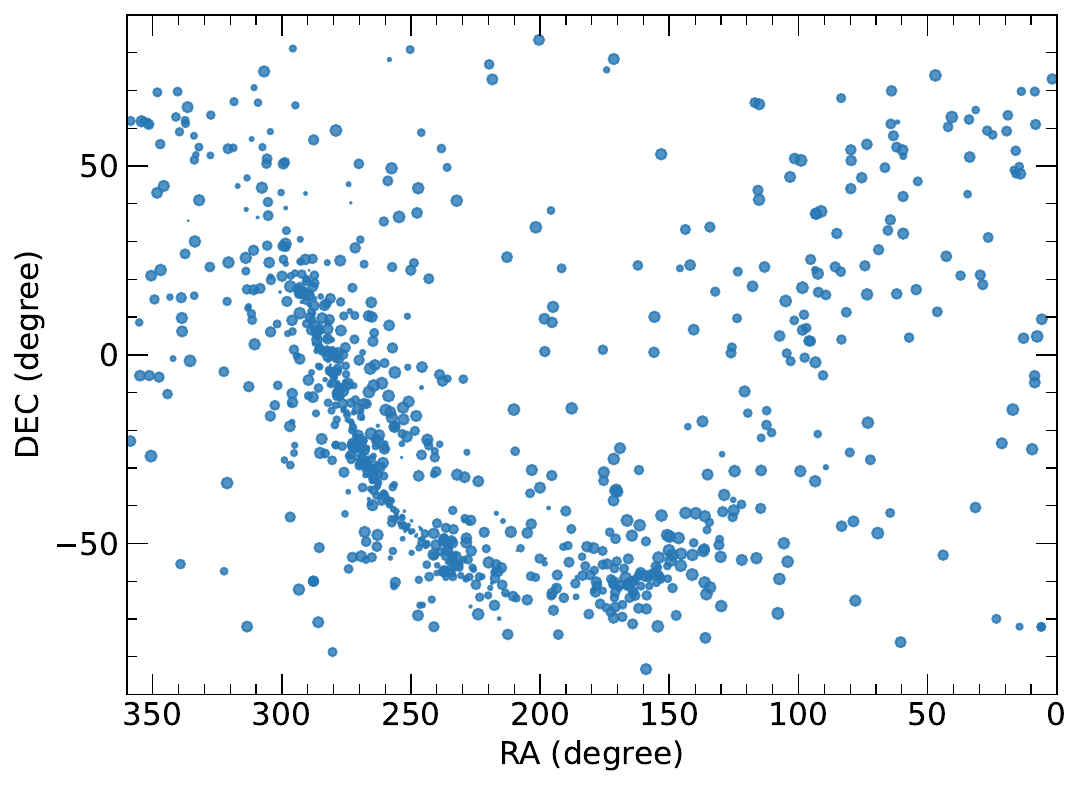}
\caption{
Locations of the 1000 pulsars we select from ATNF Pulsar Catalogue within $5$\,kpc. The size of the circles represent the relative pulsar distance with larger circles corresponding to closer pulsars. 
}
\label{fig:f1}
\end{figure}

Finally the SNR of a GW signal can be obtained by
\begin{equation}
\varrho^{2}=\sum_{j=1}^{N_{\mathrm{p}}} \sum_{i=1}^{N}\left[\frac{s_{j}\left(t_{i}\right)}{\sigma_{t, j}}\right]^{2},
\label{eq:SNR}
\end{equation}
where $N_{\rm p}$ is the total number of MSPs applied in a specific PTA experiment, $N$ is the total number of timing data points, and $\sigma_{t, j}$ is the rms of the timing precision for j-th MSP. In this paper, we consider four different PTA configurations, including two CPTA and two SKA-PTA possible configurations, see Table~\ref{tab:event_rate_1} for the detail settings. As for the pulsars location ($\alpha_{\rm p},\delta_{\rm p}$) and pulsar distance $D_{\rm p}$, we randomly select $N_{\rm p}$ pulsars with distance $D_{\rm p}<5\,\rm{kpc}$ from the the Australia Telescope National Facility (ATNF) Pulsar Catalogue\footnote{\url{https://www.atnf.csiro.au/research/pulsar/psrcat/}} and assume that they all have the same timing precision $\sigma_t$. Uniform detection cadence $\Delta t$ is assumed during the observation time. Figure~\ref{fig:f1} shows the locations of $1000$ pulsars in the sky map, in which the size of the circles stand for the relative distance of these pulsars. Thus the SNR of the GW radiates from each SMBBH in our sample can be obtained via Equation~\eqref{eq:SNR} for different PTA configurations, and further selected as ``detectable'' if its SNR is larger than a given threshold $\varrho_0$. In our following analysis, we set three different $\varrho_0$, i.e., $3$, $5$, or $10$. 

\section{Wave effect of gravitational lensing}
\label{sec:3}

\subsection{Wave effect}

\begin{figure}
\includegraphics[width=\columnwidth]{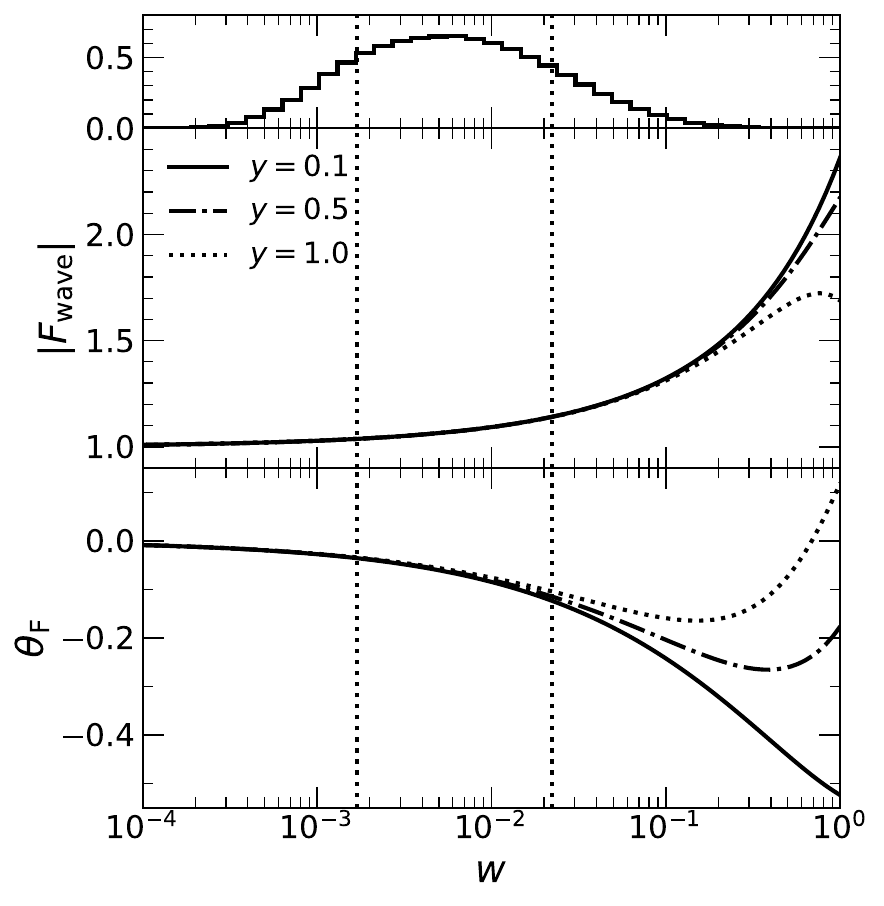}
\caption{
The amplitude $|F_{\rm wave}|$ (middle panel) and phase $\theta_{\rm F}$ (bottom panel) of the  amplification factor as a function of the dimensionless frequency parameter $w$. Solid, dash-dotted, and dot lines represent those cases with different source location in the lens plane, i.e., $y=0.1$, $0.5$, and $1.0$, respectively. Top panel shows the probability distribution of $w$ for those mock lensed SMBBHs. The vertical dotted lines stand for the 16\%-84\% range for the $w$ distribution.
}
\label{fig:f2}
\end{figure}

The amplification factor $F_{\rm{wave}}(f)$ is a complex number and fluctuate as a function of frequency due to the wave effect (diffraction and interference) \citep{Bontz=1981}. Under the thin-lens assumption, the amplification factor in the wave optics regime is give by \citep{Schneider=1992,Nakamura=1999,Takahashi=2003}
\begin{equation}
F_{\rm{wave}}(f)=\frac{D_{\rm s} \xi_{0}^{2}}{D_{\rm l} D_{\rm{ls}}} \frac{f\left(1+z_{\rm l}\right)}{i} \int d^{2} \boldsymbol{x} \exp \left[2 \pi i f T_{\rm d}(\boldsymbol{x}, \boldsymbol{y})\right],
\end{equation}
where $D_{\rm s}$, $D_{\rm l}$, and $D_{\rm ls}$ are the angular diameter distances from the source and lens to the observer, and that between the source and the lens, $f$ is the observed GW frequency at the observer's frame. Normalization constant of length $\xi_{0}$ is set to be the Einstein radius in the SIS lens model, i.e., $\xi_{0}=4 \pi (\sigma_{\rm v}/c)^{2} D_{\rm l} D_{\rm ls} / D_{\rm s}$, where $\sigma_{\rm v}$ is the velocity dispersion. Factor $(1+z_{\rm l})$ accounts for the time dilation, where $z_{\rm l}$ is the redshift of the lens. The arrival time delay through the deflected path of the propagation $T_{\rm d}(\boldsymbol{x}, \boldsymbol{y})$ is given by
\begin{equation}
T_{\rm d}(\boldsymbol{x}, \boldsymbol{y})=\frac{D_{\rm s} \xi_{0}^{2}}{D_{\rm l} D_{\rm l s}}\left(1+z_{\rm l}\right)\left[\frac{1}{2}|\boldsymbol{x}-\boldsymbol{y}|^{2}-\psi(\boldsymbol{x})+\phi_{\rm m}(\boldsymbol{y})\right],
\end{equation}
where $\boldsymbol{x}$ is the dimensionless impact parameter in the lens plane and $ \boldsymbol{y}$ is the dimensionless source location in the source plane; $\psi(\boldsymbol{x})$ is the deflection potential and $\phi_{\rm m}$ is added to set the minimal value of $T_{\rm d}(\boldsymbol{x}, \boldsymbol{y})$ as 0.

Under the assumption of the SIS lens model, $(\boldsymbol{x}, \boldsymbol{y})$ reduce to $(x,y)$ and $F_{\rm{wave}}(f)$ reduce to
\begin{equation}
F_{\rm{wave}}(w)=-i w e^{i w y^{2} / 2} \int_{0}^{\infty} d x\, x J_{0}(w x y) e^{i w\left[\frac{1}{2} x^{2}-x+\phi_{\rm m}(y)\right]}
\end{equation}
where $J_0$ is the Bessel function of the zeroth order, $w = 8\pi M_{\rm Lz} f$ is the dimensionless GW frequency, in which $M_{\rm Lz}$ is the lens mass included in the Einstein radius defined by $M_{\rm Lz}= 4\pi^{2} \sigma_{\rm v}^4 (1+z_{\rm l}) D_{\rm l} D_{\rm ls}/(G c^{2} D_{\rm s})$. Here we set $\phi_{\rm m}(y)=y+\frac{1}{2}$, which makes the minimum value of the arrival time delay to be zero.
\citet{Guo=2020} presented explicit details of multiple methods to compute the diffraction integral above. Here we adopt the $10$th-order asymptotic expansion method to calculate $F_{\rm{wave}}(f)$. By setting $z = x^2/2$, then
\begin{equation}
\begin{aligned}
F_{\rm{wave}}(w) = &C \cdot \int_{0}^{\infty} d z\, e^{i w z} g(z)\\
=& C \cdot \left[\int_{0}^{b} d z\, e^{i w z} g(z) + e^{iwb} \sum_{n=1}^{10} \frac{(-1)^{n}}{(iw)^{n}} \frac{\partial^{n-1} g(z)}{\partial z^{n-1}} \Bigg|_{z = b} \right]
\end{aligned}
\label{eq:AE}
\end{equation}
where
\begin{equation}
C = -iwe^{iw(\frac{y^2}{2}+y+\frac{1}{2})}
\end{equation}
\begin{equation}
g(z)=  e^{-iw\sqrt{2z}} J_{0}(wy\sqrt{2z})
\end{equation}
with larger $b$, Equation~\eqref{eq:AE} may relatively be more accurate and converge faster, but need longer time for computation at the same time. According to \citet{Guo=2020}, $b=10^{4}$ is a suitable value of trade-off between accuracy and computation time. Thus this value is adopted in our calculation of $F_{\rm wave}$. 

\begin{figure}
\includegraphics[width=\columnwidth]{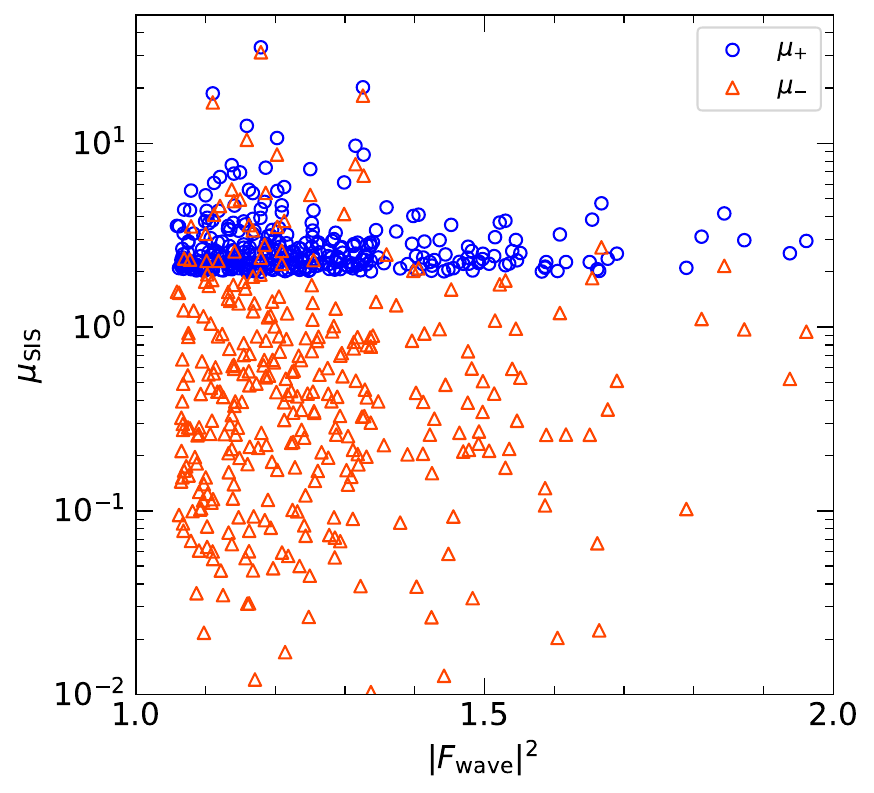}
\caption{
Comparison of the amplification factor $|F_{\rm wave}|$ obtained in the wave optics regime and the magnification factor $\mu_{\rm SIS}$ ($\mu_+$ and $\mu_-$) obtained in the geometrical optics regime for the mock lensed SMBBHs. These SMBBHs are obtained from a realization of the lensed GW events detected by SKA-opt with $\varrho \geq3$ via 30-year observations (see Tab.~\ref{tab:event_rate_1}).
}
\label{fig:f3}
\end{figure}

Figure~\ref{fig:f2} shows the amplitude $|F_{\rm wave}|$ (middle panel) and phase $\theta_{\rm F}$  (bottom panel) of amplification factor as a function of the dimensionless frequency parameter $w$, as well as the distribution of $w$ for mock lensed cosmic SMBBHs. As seen from this figure, majority of the lensed GW SMBBH sources detected in the PTA band have small $w$, e.g., $\lesssim 0.2$, and peak around $w \sim 6 \times 10^{-3}$. The lensed GW signals of most sources can only be amplified by a factor of $|F_{\rm wave}|$, which is $\sim 1.04-1.14 $ for mock BBHs with $|F_{\rm wave}|$ ranking from low to high in the $16\%-84\%$ range. Few of the lensed GW sources can be amplified by a factor of $> 1.5$. The bottom panel of Figure~\ref{fig:f2} shows the phase $\theta_{\rm F}$ of the amplification factor, suggesting the phase shift of the waveform by the gravitational lensing is typically in the range from $\sim -0.04$ to $-0.12$\,rad (corresponding to the $16$\% to $84$\% range of $\theta_{\rm F}$ with ranking from low to high).

Both the amplitude and the phase of amplification factor for most lensed nHz GW signals are insensitive to the source location due to small $w$ ($\lesssim 0.2$). This is totally different from the case in the geometrical optics regime, for which the magnification factor highly depends on the source location $y$, e.g., $\mu_{+} = (1+y)/y$ and $\mu_{-} = (1-y)/y$ for the SIS lens model. Note here that, in accordance with previous papers, we use the term $\mu$ to describe the amplification of lensed images in the geometrical optics regime, and use the term $F_{\rm wave}(f)$ to describe the amplification in the wave optics regime. For $w \gtrsim 1$ where $F_{\rm wave}(f)$ asymptotically converges to the geometrical optics approximation, $F_{\rm wave}(f)$ and $\mu$ are related by $|F_{\rm wave}(f)|^{2} = |\mu_{+}| + |\mu_{-}| + 2|\mu_{+}\mu_{-}|^{1/2} \sin(2 \pi f T_{\rm d})$ for $y < 1$ and $\mu_{-} =0$ for $y \geq 1$. Figure~\ref{fig:f3} shows the magnification factor for those mock lensed sources in the geometrical optics regime (blue symbol for $\mu_+$ and orange symbol for $\mu_-$) and the amplification factor in the wave optics regime ($|F_{\rm wave}(f)|^{2}$), respectively, by assuming the SIS model for the lens. Apparently, the magnification effect resulting from the wave optics is substantially different from that resulting from geometrical optics. In the wave optics regime, $|F_{\rm wave}(f)|^{2}$ is in the range of $\sim 1-2$, while in the geometrical optics regime, $\mu_{+}$ and $\mu_-$ are in the range of $\sim 2-40$ and $\sim 10^{-2}-30$, respectively. For the same lensed source, $\mu_+$ is always larger than $|F_{\rm wave}(f)|^{2}$ and can reach higher magnification factor up to several tens for some extreme cases near the caustic, and $\mu_-$ can be larger than $1$ if $y<0.5$ and smaller than $1$ if $1> y>0.5$ (the lensed source is demagnified), while $|F_{\rm wave}(f)|^{2}$ is always slightly larger than 1. These prove again the importance to take the wave effect into consideration for the detection of lensed nHz GW signals from SMBBHs.

\subsection{Lensed waveforms}
\label{sec:waveform}

\begin{figure*}
\includegraphics[width=0.9\textwidth]{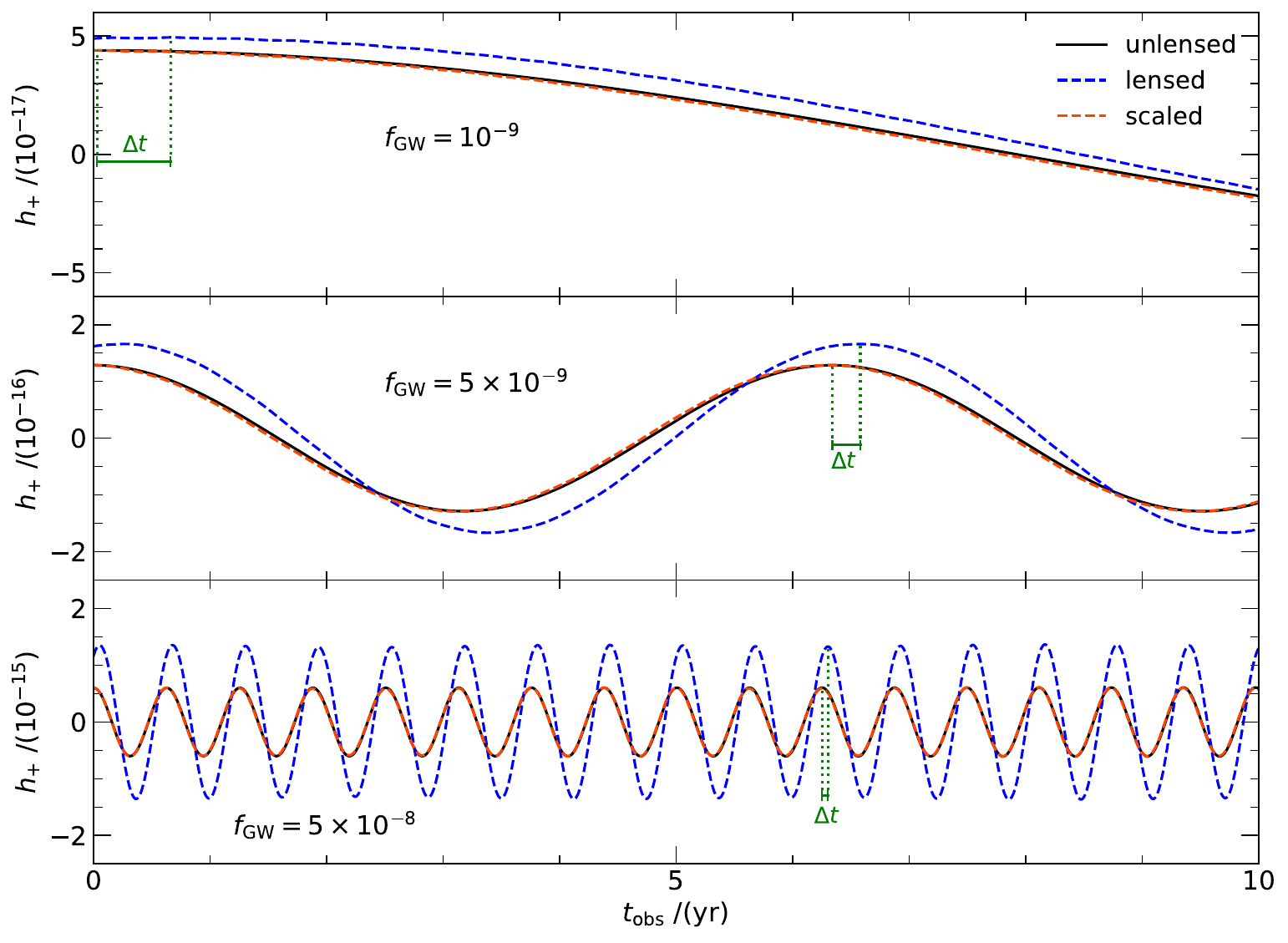}
\caption{
The waveforms of GWs with/without considering lensing effect emitted from example SMBBHs at redshift $z_{\rm s}=1$ during 10 years observation. Top, middle, and bottom panels show the cases for SMBBHs with component masses of $m_1=m_2=10^9M_\odot$, emitting GWs at frequency in the observer's frame $f_{\rm GW,0}=10^{-9}$, $5 \times 10^{-9}$, and $5 \times 10^{-8}$\,Hz, respectively.
In each panel, black solid, blue dashed, and orange dashed lines show the GW waveforms from the SMBBH without lensing (``unlensed''), with lensing (``lensed''), and the one (``scaled'') by scaling the amplitude of the lensed one by a constant factor ($1.12$/$1.29$/$2.25$) and a phase shift ($\Delta \theta_{\rm F}=2\pi\Delta t/T_{\rm P}$=$0.13$/$0.27$/$0.52$\,rad with $\Delta t$ representing the observation time shift between the peaks of the lensed GW signal and the original one as marked in each panel and $T_{\rm P}$ representing the period of the system in the observer's frame), respectively.
For simplicity, the inclination angle $\iota$ and the polarization angle of GW $\psi$ are set to be 0, the source location is fixed at $y=0.1$ where the lens locates at $z_{\rm l}=0.5$, and we use $h_+$ to represent the GW waveform.
}
\label{fig:f4}
\end{figure*}

For the GW signal from an inspiralling SMBBH on circular orbit, its waveform in the time domain is given by Equations~\eqref{eq:h_p} and \eqref{eq:h_t}. In order to find out how the GW waveform is changed by the gravitational lensing in the wave optics regime, we calculate some typical GW waveforms and the corresponding lensed waveforms. The original waveform without lensing is represented by $h_+$ for simplicity (the real waveform is actually a combination of $h_+$ and $h_\times$). As for the lensed GW waveform, it highly depends on the GW frequency and the total observation duration as shown in Figure~\ref{fig:f2}. First, we obtain the GW waveform in the frequency domain which is Fourier transformed from the original GW waveform in the time domain. Then we magnify this transformed waveform by $F_{\rm wave}(w)$ according to its frequency. Finally, we obtain the lensed waveform in the time domain by taking the inverse Fourier transform for the magnified waveform in the frequency domain. 

Figure~\ref{fig:f4} shows the lensed GW waveform for three example SMBBHs in the time domain. For these SMBBHs, we set equal component mass with $m_{1}=m_{2}=10^{9} M_{\odot}$, redshift $z=1$,  inclination angle $\iota = 0$, polarization angle of GW $\psi = 0$, source location fixed at $y=0.1$ in the lens plane, lens at $z_{\rm l}=0.5$, but different initial GW frequency $f_{\rm GW,0}=10^{-9}$\,Hz (top panel), $5 \times 10^{-9}$\,Hz (middle panel), and $5 \times 10^{-8}$\,Hz (bottom panel), respectively. The original and lensed waveforms for these three cases are shown in black solid and blue dashed lines, respectively. The orange dashed lines are the scaled lensed waveforms, where we match the lensed and original waveforms at the initial point. Apparently, all the waveforms are nearly sinusoidal, the scaled lensed waveforms can match the original waveforms almost exactly, as the frequency evolution within 10\,yr observation is insignificant and can be neglected. This indicates that the lensing information may be degenerated with the SMBBH parameters (e.g., $m_{1},m_{2},z,\iota,\psi$).

This feature of nHz GW signals is straightforward to understand. Unlike those high frequency GW signals (e.g., GWs emitted from mergers of compact binaries) which have rapid frequency evolution during the observation duration ($\lesssim$100\,s), the typical frequency of GW signals detected by PTA is in the range of $10^{-9}-10^{-7}$\,Hz and the observation duration lasts about a decade or even a few decades. This means that the frequency evolution within the observation duration is tiny according to Equation~\eqref{eq:f_GW_evolution}, thus the GW signal is nearly monochromatic during the whole observation period. For instance, the period of a GW signal emitted by SMBBH with initial frequency $f_{\rm GW,0}=10^{-8}$\,Hz is $\sim 3$yr. The relative frequency difference during a ten-year observation is only $|(f_{\rm GW,0}-f_{\rm GW,10})/f_{\rm GW,0}| \sim 10^{-6}$, thus $f_{\rm GW}$ almost remains unchanged during the entire ten year observation (i.e., three times the GW waveform period).

We note here that some lensed SMBBHs may be highly eccentric in reality, for which the GW signals emitted at a number of different harmonics may have comparable power and be detectable, and thus their GW signals are not monochromatic as that in the cases with circular orbits. The amplification factors for GW signals from different harmonics can be substantially different. Therefore, it is possible to identify the lensed events via the GW signals from SMBBHs with large eccentricities. We defer detailed investigation on the lensing effect of GW signals from eccentric SMBBHs to a future work.

\section{Lensed SMBBHs in the PTA band}
\label{sec:4}

\newcommand{\tabincell}[2]{\begin{tabular}{@{}#1@{}}#2\end{tabular}}

\begin{table*}
\caption{Estimates for the total number of detectable lensed nHz GW signals and their associated host galaxy detected by different PTA experiments (CPTA and SKA-PTA) for different SNR thresholds $\varrho_{0}=3/5/10$.}
\begin{adjustbox}{center}
\begin{threeparttable}
\renewcommand\arraystretch{1.2}
\setlength\tabcolsep{3pt}
\begin{tabular}{lcccccccccc}		\toprule
PTA    &$\varrho_{0}$ & $N_{\rm p}$& \tabincell{c}{$\sigma_{t}$ \\ (ns)} &\tabincell{c}{$\Delta t$\\ (week)} &  \tabincell{c}{Detectable \\ GW (10\,yr)} &  \tabincell{c}{Lensed GW\\ (10\,yr)} & \tabincell{c}{Lensed Host\\ (10\,yr)} &  \tabincell{c}{Detectable \\ GW (30\,yr)}&  \tabincell{c}{Lensed GW\\ (30\,yr)} & \tabincell{c}{Lensed Host\\ (30\,yr)}    \\ \toprule
\multirow{3}{*}{$\textbf{CPTA}$}
&3  & 100 &100 &2  & 86   &0.00    &0.00   & $1.91$$\times$$10^{4}$   &1.89  &0.28\\
&5  & 100 &100 &2  & 21   &0.00    &0.00   & $5.60$$\times$$10^{3}$   &0.64  &0.07\\
&10 & 100 &100 &2  & 3    &0.00    &0.00   & $9.00$$\times$$10^{2}$   &0.09  &0.02\\ 
\midrule
\multirow{3}{*}{$\textbf{CPTA-opt}$}
&3  & 100 &20 &1  & $7.74$$\times$$10^{3}$   &1.01   &0.12    & $9.54$$\times$$10^{5}$  &106     &14.9\\
&5  & 100 &20 &1  & $2.60$$\times$$10^{3}$   &0.37   &0.06    &  $3.81$$\times$$10^{5}$ &44.8    &6.31\\
&10 & 100 &20 &1  & $5.45$$\times$$10^{2}$   &0.09   &0.00   &  $9.24$$\times$$10^{4}$ &10.9    &1.65\\ \midrule
\multirow{3}{*}{$\textbf{SKA}$}
&3  & 1000 &100 &2  & $1.41$$\times$$10^{3}$   &0.26    &0.03        & $2.12$$\times$$10^{5}$  &25.3    &3.67\\
&5  & 1000 &100 &2  & $4.27$$\times$$10^{2}$   &0.08    & 0.00       & $7.45$$\times$$10^{4}$  &8.52    & 1.26\\
&10 & 1000 &100 &2  & 90                       &0.01     & 0.00      & $1.60$$\times$$10^{4}$   &1.76     &0.20\\ \midrule
\multirow{3}{*}{$\textbf{SKA-opt}$}
&3  & 1000 &20 &1  & $6.44$$\times$$10^{4}$  &7.44     &1.15    & $3.20$$\times$$10^{6}$  &289     &37.4\\
&5  & 1000 &20 &1  & $2.68$$\times$$10^{4}$  &3.60     &0.56    & $2.24$$\times$$10^{6}$  &219     &29.4\\
&10 & 1000 &20 &1  & $7.02$$\times$$10^{3}$  &0.95     &0.12    & $8.75$$\times$$10^{5}$  &97.7     &13.5\\  \bottomrule
\end{tabular}
\begin{tablenotes}
\footnotesize
\item \textit{Note.} 3rd, 4th, and 5th columns list the detailed configuration of different PTA experiments including the millisecond pulsar number (`$N_{\rm p}$'), the pulsar timing precision (`$\sigma_{t} $'), and the detection cadence (`$\Delta t $'). The following three columns are the total event number for the detectable nHz GW signals without considering gravitational lensing (`Detectable GW'), the detectable lensed nHz GW signals (`Lensed GW'), and the detectable associated host galaxy (`Lensed Host'), respectively, assuming the total observation time as 10\,yr. The last three columns are the same but with the total observation time as 30\,yr. For 10-year detection, we cut the GW frequency at $1/T_{\rm obs}$, where $T_{\rm obs}$ is the observation time.
\end{tablenotes}
\end{threeparttable}
\end{adjustbox}
\label{tab:event_rate_1}
\end{table*}
Each SMBBH in our sample obtained in Section~\ref{sec:SMBBH_sample} may have a chance to be strongly lensed by a foreground galaxy with a probability characterized by the \textit{optical depth} $\tau$. Assuming the SIS model, this optical depth for a source at redshift $z_{\rm s}$ is given by
\begin{eqnarray}
\tau_{\rm SIS} (z_{\rm s}) & = & \int_{0}^{z_{\rm s}} \frac{dV}{dz_{\rm l}} dz_{\rm l} \int^{\infty}_{0}  \frac{d N}{d \sigma_{\rm v}} \frac{A_{\rm SIS}}{4\pi} d \sigma_{\rm v}, \nonumber \\
& = &
\int^{z_{\rm s}}_0 dz_{\rm l} \int^\infty_0 d\sigma_{\rm v} P(z_{\rm l},\sigma_{\rm v}|z_{\rm s}).
\label{eq:tau_sis}
\end{eqnarray}
In the above Equation, $dV/dz_{\rm l}$ is the comoving volume element per unit redshift and it is equal to $4\pi c D_{\rm c,l}^2 H_{0}^{-1} [\Omega_{\rm m}(1+z_{\rm l})^3+\Omega_{\Lambda}]^{-1/2}$ for the adopted $\Lambda$CDM model, with $D_{\rm c,l}$ denoting the comoving distance to the lens. The cross-section for lensing to occur $A_{\rm SIS}$ is equal to $\pi y^2_{\rm m} \theta_{\rm E}^{2}$ for the SIS model, with $\theta_{\rm E} = 4\pi(\sigma_{\rm v}/c)^{2}(D_{\rm ls}/D_{\rm s})$ denoting the angular Einstein radius for the SIS model, and $y_{\rm m}$ is the boundary of the cross-section. $dN/d\sigma_{\rm v}$ is the distribution of the velocity dispersion for SIS lenses. The probability for a GW event at $z_{\rm s}$ is lensed is $p(z_{\rm s}) = 1- \exp\left(-\tau_{\rm SIS}(z_{\rm s})\right) \simeq \tau_{\rm SIS}(z_{\rm s})$ as $\tau_{\rm SIS} \ll 1$. The term $P(z_{\rm l},\sigma_{\rm v}|z_{\rm s})$ $\left(= \frac{dV}{dz_{\rm l}} \frac{d N}{d \sigma_{\rm v}} \frac{A_{\rm SIS}}{4 \pi} \propto \frac{d N}{d \sigma_{\rm v}} \theta_{\rm E}^{2} D_{\rm c,l}^2 [\Omega_{M}(1+z_{\rm l})^3+\Omega_{\Lambda}]^{-1/2}\right)$ in the second line of Equation~\eqref{eq:tau_sis} denotes the probability distribution of a GW event at redshift $z_{\rm s}$ that is lensed by a galaxy with velocity dispersion in the range of $\sigma_{\rm v}\rightarrow \sigma_{\rm v}+d\sigma_{\rm v}$ at the redshift range of $z_{\rm l}\rightarrow z_{\rm l}+ dz_{\rm l}$.

We only consider those systems with $y\leq1$ as the ``strongly'' lensed ones in this paper. In principle, systems with $y$ slightly larger than $1$ can also have lensing effect in the wave optics regime but less significant than those with $y<1$. For the lensing of the SMBBH host galaxies (in the geometrical optics regime), it is in the strong lensing regime when $y<1$ and in the weak lensing regime when $y>1$. For simplicity, we do not consider those cases with $y>1$ in the following analysis and thus the cross-section reduces to $A_{\rm SIS} = \pi \theta_{\rm E}^2$.

The galaxy-scale strong lensing we considered here is mainly caused by the intervening early-type galaxies \citep{Turner=1984,Moller=2007}.\footnote{Massive lenes, such as galaxy groups or clusters, may lead to more significant lensing effects on both the GW signals and their host galaxies due to the larger amplification caused by them \citep{Smith=2018,LIGO2021lensing}. The lensing events caused by these massive lenses with relatively higher amplification may be easier to be observed by EM observations as they are rare compared with elliptical galaxies, which deserves further investigation. For simplicity, we do not consider them in the present work.} 
The number density distribution of these galaxies $dN/d\sigma_{\rm v}$ can be described by a modified Schechter function \citep{Choi=2007} and the redshift evolution of this number density distribution may be simply described by the power-law evolution model, i.e., 
\begin{equation}
\frac{\mathrm{d} N}{\mathrm{~d} \sigma_{\rm v}}=\phi_{z} \left(\frac{\sigma_{\rm v}}{\sigma_{z} }\right)^{\alpha} \exp \left[-\left(\frac{\sigma_{\rm v}}{\sigma_{z}}\right)^{\beta}\right] \frac{\beta}{\Gamma(\alpha / \beta)} \frac{1}{\sigma_{\rm v}},
\label{eq:sigma_1}
\end{equation}
and
\begin{equation}
\phi_{z} = \phi_{*} \left( 1+z_{\rm l} \right)^{\kappa_n};\ \ \sigma_{z} = \sigma_{*} \left( 1+z_{\rm l} \right)^{\kappa_{\rm v}}.
\label{eq:sigma_2}
\end{equation}
Here ($\phi_{*}$, $\sigma_{*}$, $\alpha$, $\beta$)=($8.0 \times 10^{-3}h^{3} {\rm Mpc}^{-3}$, $161\ {\rm km\ s^{-1}}$, $2.32$, $2.67$) according to the fitting results given by \citet{Choi=2007}, and the two redshift evolution parameters $\kappa_n=-1.18$ and $\kappa_{\rm v}=0.18$ according to \citet{Geng=2021}.

We adopt the Monte-Carlo method to generate mock lensed SMBBHs according to the sample of cosmic SMBBHs given in Section~\ref{sec:SMBBH_sample} and the distribution $P(z_{\rm l},\sigma_{\rm v}|z_{\rm s})$. Then we calculate the GW signals emitted from these SMBBHs and estimate their expected SNR for any given PTA configuration as those listed in Table~\ref{tab:event_rate_1}. As the lensed SMBBHs radiate continuous GW signals which can be approximately assumed as monochromatic, their SNRs can be directly estimated as $\varrho_{\rm l}  = |F_{\rm wave}(f_{\rm GW})| \varrho$, in which $\varrho$ represents the SNR of the original GW signals without lensing. We set several different threshold values as $\varrho_0=3/5/10$ to define a source (either the lensed one or the one without lensing) as ``detectable''. These SNR limited samples contain some GW sources with $\varrho<\varrho_0$ and $\varrho_{\rm l}>\varrho_0$,  which are detectable because of the lensing magnification bias. Note that this magnification bias is small for the cases considered here due to the small amplification factor $|F_{\rm wave}(f_{\rm GW})|$. In addition, the lowest observable GW frequency is set to $10^{-9}$\,Hz, which corresponds to a waveform period up to $\sim 30$\,years. Thus we consider two scenarios of observation time $T_{\rm obs}$ as 10 years and 30 years. For the 10-year observation, we set a lower limit for detectable GW signals as $1/T_{\rm obs}$ to ensure at least a whole waveform can be observed. In order to obtain the total number of detectable lensed SMBBHs by any PTA configuration for different scenarios, we generate 100 independent such realizations and adopt their average detectable lensed number as our results.

\begin{figure*}
\centering
\includegraphics[width=2\columnwidth]{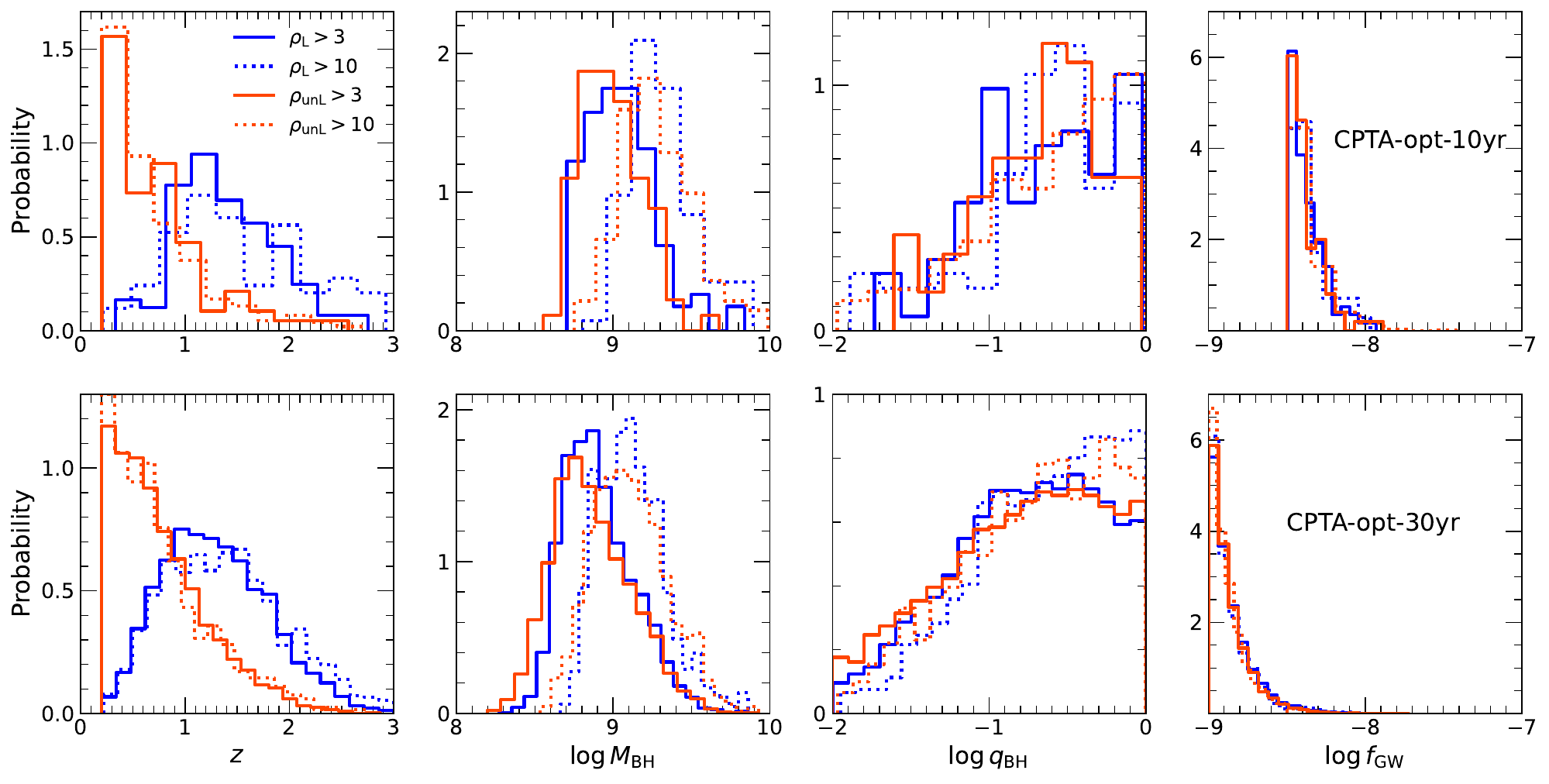}
\caption{
Probability distributions of the parameters for the mock SMBBHs that can be detected by CPTA with the optimistic configuration. Panels from left to right show the distributions of SMBBH redshift $z$, total mass $M_{\rm BH}$, mass ratio $q_{\rm BH}$, and GW frequency $f_{\rm GW}$, respectively. Blue and orange lines show the distributions for mock SMBBHs with and without lensing, with SNR larger than $3$ (solid lines) or $10$ (dotted lines), respectively. Top and bottom panels show the distributions for SMBBHs detected via $10$ and $30$ years observations, respectively.
}
\label{fig:f5}
\end{figure*}
\begin{figure*}
\centering
\includegraphics[width=2\columnwidth]{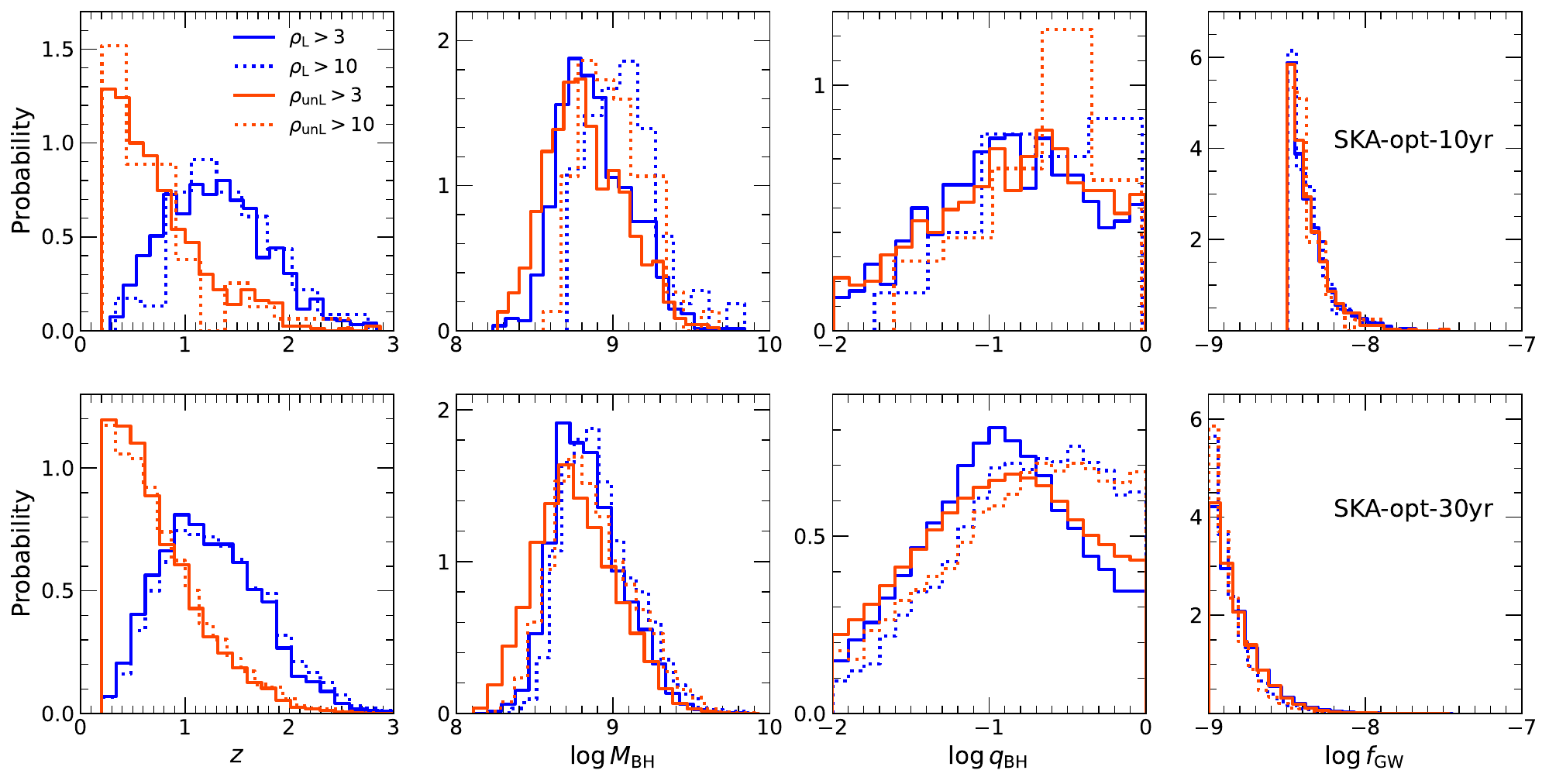}
\caption{
Legend is the same as that for Fig.~\ref{fig:f5}, except that the SMBBHs are detected by SKA-PTA with optimistic configuration.
}
\label{fig:f6}
\end{figure*}

Table~\ref{tab:event_rate_1} lists the expected number of detectable SMBBHs, either the lensed cases or the cases without lensing, for future PTA experiments with different configurations.
For observation period as $10$\,years, future optimal PTA configurations, such as CPTA-opt and SKA-opt, may be able to detect up to a large number of SMBBHs (more than a few thousands or tens of thousands if $\varrho_0=10$ or $3$; see Table~\ref{tab:event_rate_1}), much higher than that detected by conservative CPTA and SKA configurations (several or several dozen for CPTA and hundreds or one thousand for SKA if $\varrho_0=10$ or $3$). However, if the observation time is extended to 30 years, $\sim 10^{3}-10^{6}$ GW signals can be detected by any future CPTA and SKA-PTA configurations. Though the expected number of detection is promising and encouraging, one should be cautious that it may be extremely challenge to subtract the GW signals of these individual sources from the PTA data. There may be GW signals from many sources overlapped together in each frequency bin, especially at low frequencies. Resolving them from the GW background could be an important task in future PTA data analysis. Among those detectable systems, only a fraction of $\sim 0.01\%$ can be strongly lensed by foreground galaxies. For an observation period of $10$\,years, both conservation configurations of CPTA and SKA can hardly detect even a single lensed nHz GW signal. Meanwhile, about $1.01/0.09$ and $7.44/0.95$ can be detected by CPTA-opt and SKA-opt assuming $\varrho_0=3/10$. As for an observation period of 30 years, the detectable lensed numbers increase accordingly, and about $1.89/0.09$ and $25.3/1.76$ can be detected by two conservation configurations of CPTA and SKA, $106/10.9$ and $289/97.7$ detected by CPTA-opt and SKA-opt assuming the same SNR threshold $\varrho_0=3/10$. Thus, within our total sample future PTA observations may be able to detect up to several tens to hundreds lensed SMBBHs as individual GW sources. 

Summing up all the 100 realizations, we can further obtain the parameter distributions (i.e., $z, M_{\rm BH}$, $q_{\rm BH}$, $f_{\rm GW}$) for those detectable SMBBHs. Figure~\ref{fig:f5} presents these distributions for the lensed SMBBHs (blue lines) and the SMBBHs without lensing (orange lines) detected by future CPTA-opt assuming SNR threshold as 3/10 (solid/dotted lines). Upper and lower panels show the different scenarios of observation time as 10 and 30 years. Similar to Figure~\ref{fig:f5}, Figure~\ref{fig:f6} shows the parameter distributions for those SMBBH with/without lensing detected by future SKA-opt. As seen from these two figure, the majority of the detectable lensed SMBBHs are distributed at redshift $z_{\rm s}\sim 0.5-2$ and peak at $z_{\rm s} \sim 1$, while those without lensing are distributed toward lower redshift and peaked at $z_{\rm s} \sim 0.2$. The difference between these redshift distributions is simply due to that sources at higher redshift have significantly larger optical depth. In general, the lensed SMBBHs have slightly larger total mass than the SMBBHs without lensing, both of which are in the range of $M_{\rm BH} \sim 10^{8}-10^{10} M_{\odot}$. For different SNR threshold $\varrho_0=3$/$10$, the total mass of lensed SMBBHs peak at $\log M_{\rm BH} \sim 8.8/9.2$. Similarly, the lensed SMBBHs and the SMBBHs without lensing have mass ratios peaked at $q_{\rm BH} \sim 0.1/0.3$ if set $\varrho_0=3$/$10$. In general, the total mass and mass ratio for detectable SMBBHs with $\varrho>3$ are smaller than those with $\varrho>10$. As for the GW frequency distribution, most the SMBBHs have frequency of detectable GW signals clustered around $10^{-9}$\,Hz ($\approx 1/30$\,yr) for 30-year observation and $3.17\times 10^{-9}$\,Hz ($\approx 1/10$\,yr) for 10-year observation. For both scenarios, only a small fraction of the detectable GW signals have $f_{\rm GW} >10^{-8}$\,Hz ($\lesssim 3\%$ for 10-year observation and $\lesssim 0.2\%$ for 30-year observation).

\begin{figure}
\centering
\includegraphics[width=\columnwidth]{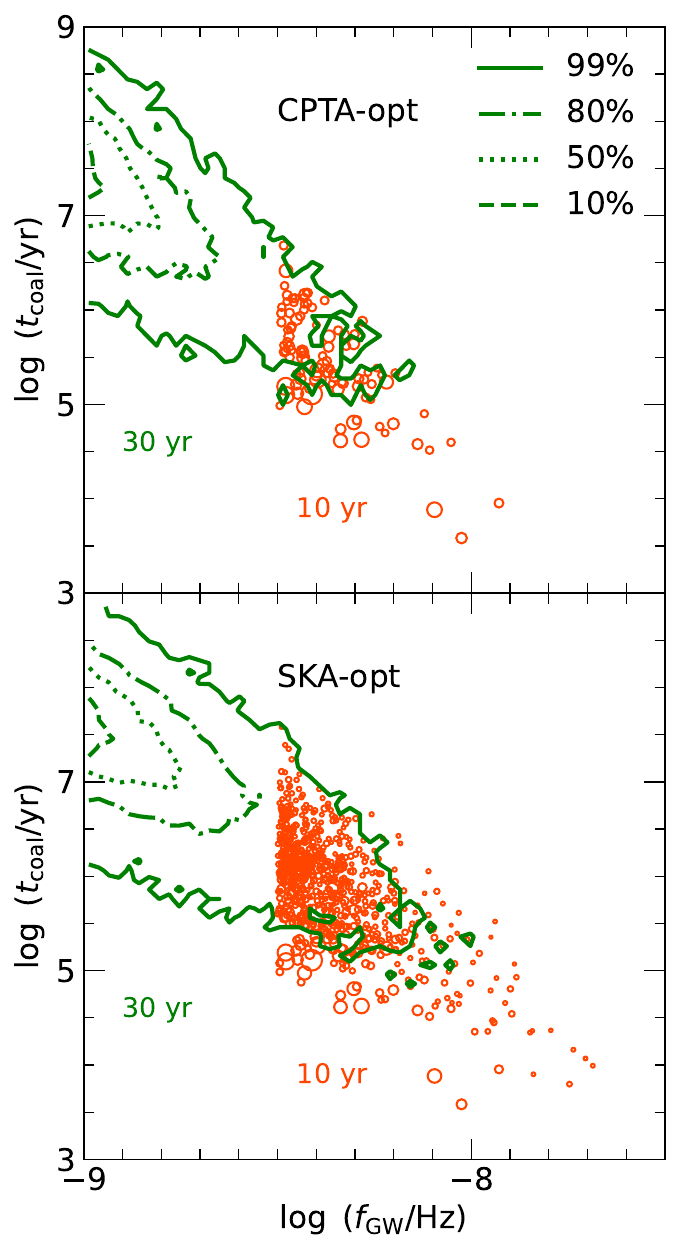}
\caption{
Distributions of the SMBBH coalescence time $t_{\rm coal}$ (in the unit of yr) and its GW frequency $f_{\rm GW}$ for the lensed sample with the SNR threshold $\varrho_0 = 3$ detected by CPTA-opt (upper panel) and SKA-opt (lower panel). In each of the panel, green contour shows the lensed SMBBH sample with observation time as 30 years, in which four types of line stand for different confidence levels of 10/50/80/99 per cent. Orange circles stand for the distributions from the lensed SMBBH sample with observation time as 10 years and with size of circle representing the relative SNR (the larger size, the larger SNR).
}
\label{fig:f7}
\end{figure}

Figure~\ref{fig:f7} shows the distribution of coalescence time $t_{\rm coal}$ and GW frequency $f_{\rm GW}$ for those lensed SMBBHs detected by CPTA-opt (upper panel) and SKA-opt (lower panel) with SNR $\varrho \geq \varrho_0=3$ for observation time as $10$ years (orange circles) and $30$ years (green contour). The coalescence time is estimated by $t_{\rm coal} = 5/256 [\pi (1+z) f_{\rm GW}]^{-8/3} (G \mathcal{M}_{\rm c} / c^3)^{-5/3}$ \citep{Maggiore=book}. As seen from this figure, more than 80 per cent of the lensed SMBBHs detected via $10$-year or $30$-year observations have the coalescence time in the range from $\sim 10^5$\,yr to $10^7$\,yr or from $\sim 10^6$\,yr to $10^8$\,yr, which are far longer than the observation time. We note that there are some extreme cases of SMBBH inspiral with $\tau_{\rm coal}$ smaller than $10^5$\,yr or even $10^4$\,yr. For $T_{\rm obs}=10/30$\,yr there are $\sim 17\%$/$\sim 0.33\%$ lensed GW sources detected by CPTA and $\sim 7.4\%$/$\sim 0.29\%$ for the lensed GW sources detected by SKA-PTA smaller than $10^5$\,yr, while there are $\sim 3.0\%$/$\sim 0.038\%$ lensed GW sources detected by CPTA and $\sim 0.94\%$/$\sim 0.035\%$ for the lensed GW sources detected by SKA-PTA smaller than $10^4$\,yr. These extreme sources may also have relatively higher SNRs (see orange circles in Figure~\ref{fig:f7}) at the same time, and may be the first ones being detected by future PTA experiments.

\section{Discussions on identification of lensed SMBBHs detected by PTA  }
\label{sec:5}

The detection of lensed SMBBHs by future PTA experiments is promising according to the above estimates. However, it is difficult to identify the detected SMBBH as a lensed system via the GW signal alone. Unlike lensed GW signals with much higher frequency such as those from mergers of compact binaries\footnote{The lensed high frequency GW sources have multiple lensing images, and the lensed GW signals can be directly identified by matching their sky localization and lens system parameters derived from different lensing images such as time delay, magnification ratio, and phase shift \citep[e.g.,][]{Haris=2018, Hannuksela=2019, Dai=2020, LIGO2021lensing}.}, a lensed nHz GW source does not have well separated multiple images of the GW signal. Its lensed waveform is different from the original one only by a small amplitude magnification and a phase shift, which can be matched well by another unlensed nHz GW signal with slightly larger chirp mass or smaller source distance compared with the true system. However, one may still be able to identify it if its associated EM counterparts (e.g., host galaxy and/or host AGN) could be also observed by future sky surveys, with which one may be able to determine the orbital and other physical parameters of the central SMBBH. Thus the lensing signatures can be found from the EM counterparts of the systems and applied to identify corresponding lensed nHz GW signals. This approach is indeed difficult, nevertheless, there are already many efforts on finding SMBBH candidates (for a summary about the current electromagnetic observational evidence of SMBBHs, see reviews by \citealt{Tamara=2022} and \citealt{Wang=2020}). We further discuss below the detectability of the associated host galaxies for the lensed nHz GW signals as well as possible active SMBBHs, and discuss the possibility of distinguishing the lensed GW signals.

\subsection{Lensed host galaxies}

For the host galaxies in general, their morphology must be considered when investigating the gravitational lensing effect, which is different from the cases of lensed QSOs and supernovae as point sources \citep{Oguri=2010}. Therefore, one has to consider a more complicated criterion for the lensed hosts by combining brightness and morphology. Based on the SIS lens model and assuming $y<1$, the host galaxy associated with the lensed nHz GW signals will also be strongly lensed and produce extended lensed images with distorted shapes (arcs or rings for highly aligned lens system). In order to identify a lensed galaxy, the distorted image should be detectable and resolvable. In addition, for an extended source, the distorted morphology of the lensed image should also be identifiable. Therefore, we adopt the criteria of defining a detectable lensed galaxy as follows \citep{Collett=2015,ChenLu2022}: (1) the lensed host galaxy can be observed by a given telescope with limiting magnitude for extended source $m_{\rm lim}$; (2) the image and counter-image must be resolvable, i.e., $R_{\rm e}^2+(s/2)^{2} \lesssim \theta_{\rm E}^2$, where $R_{\rm e}$ is the effective radius of the galaxy and $s$ is the angular resolution; (3) the tangential shearing of the arcs should be detectable, i.e., ${\mu_{\rm tot} R_{\rm e} > s}$ and $\mu_{\rm tot} > 3$, where $\mu_{\rm tot}$ is the total magnification of the source. 

For future sky survey, one of the core community surveys for Roman Space Telescope (RST) is the High Latitude Wide Area Survey\footnote{\url{https://roman.gsfc.nasa.gov/high_latitude_wide_area_survey.html}}, which covers four near-infrared bands (F106, F129, F158, F184). We first assume that the lensed PTA sources can be surveyed by RST without sky coverage limitation, but discuss later about this limitation. For example, we choose F158 filter as the representative of RST and mainly discuss the detectability of host galaxies for RST. RST/F158 has the limiting magnitude for extended galaxy source $m_{\rm lim} \sim 26.8$ and angular resolution $s\sim 0.11$\,arcsec. Then, for those lensed GW signals in each realization obtained in the last section, corresponding effective radius $R_{\rm e}$ and the magnitude in F158 filter for each SMBBH host galaxy are obtained by sampling from JAGUAR \citep{Williams=2018}, the extragalactic mock galaxy catalog designed for James Webb Space Telescope (JWST), in different bins of redshift and galaxy mass. JAGUAR mock galaxy catalog covers the redshift range between $0.2$ to $15$, which is sufficiently deep for our SMBBH samples ($0.2<z<3$). For each mock galaxy in the catalog, the flux information in all JWST band filters and the stellar mass of the galaxy are provided. We obtain the flux in RST/F158 filter by interpolation according to the flux in the JWST bands (i.e., F150W and F162M) and then convert it to the AB magnitude in the RST/F158 filter. The total mass of SMBBH is correlated with the mass of the host galaxy $M_{\rm gal}$ through the MBH mass-bulge mass relation, i.e., $p_{\mathrm{BH}}(M_{\mathrm{BH}}, q_{\mathrm{BH}} \mid M_*, q_*,z')$ introduced in section~\ref{sec:SMBBH_sample}. 

By applying the above criteria, we can further obtain the subsample composed of detectable lensed GW signals and associated lensed host galaxies for each realization. Similarly, the detectable number of the associated lensed host galaxies is obtained through averaging all the realizations. Results are presented in the eighth column and the last column in Table~\ref{tab:event_rate_1} for the cases with a PTA observation period of $10$~years and $30$~years, respectively. We find that there are $\sim 14\%$ of the lensed nHz GW signals having detectable host galaxies passing our lensing criteria. For the cases with a PTA observation period of $10$\,years, it is hard to detect even one system with both detectable GW signals and detectable lensed host galaxy for the PTA configurations investigated above, though SKA-opt may detect $\sim 1$ such a case with SNR $\geq 3$. For the cases with a PTA observation period of $30$\,years, $\sim 0.28$ or $0.02$ and $\sim 3.67$ or $0.20$ host galaxies of those lensed GW signals can be detected by RST associated with the conservative CPTA and SKA assuming SNR threshold $\rho_0=3$ or $10$, while $\sim 14.9$ or $1.65$ and $\sim 37.4$ or $13.5$ such cases may be detected by RST associated with the CPTA-opt and SKA-opt. Thus for an unknown lensed nHz GW signal, its lensed host galaxy may be detected and further help identify the corresponding GW signals. The probability to detect the lensed host galaxies may vary with how the lensing morphology criterion is set rather than the brightness criterion for the host galaxies. When the morphology criterion (3) is changed to $\mu_{\rm tot}>4$ or $5$, which means a much sharper arcs, this probability decreases to $\sim 9\%$ or $5\%$. However, as for the brightness of the host galaxies, nearly all the host galaxies are bright enough to be observed by RST. 

\begin{figure}
\centering
\includegraphics[width=\columnwidth]{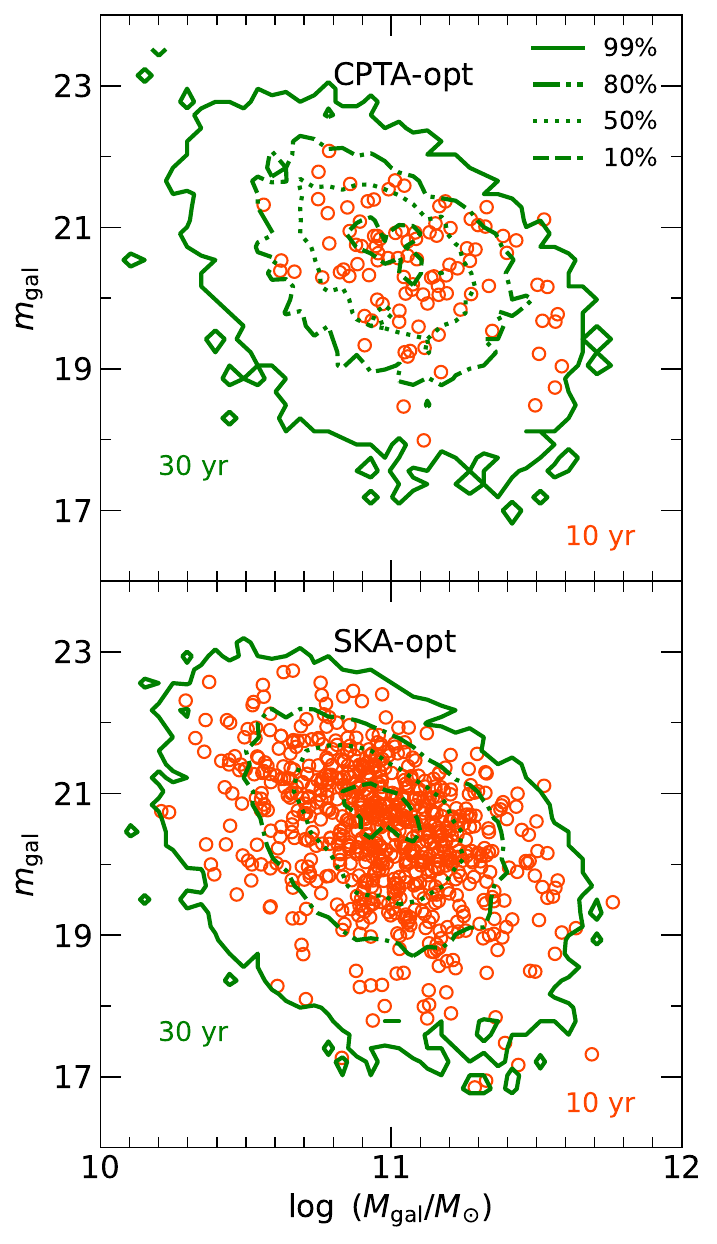}
\caption{
Distributions of the host galaxy mass and galaxy magnitude for our lensed SMBBH sample with the SNR threshold $\varrho_0 = 3$ detected by CPTA-opt (upper panel) and SKA-opt (lower panel). In each of the panel, green contour shows the lensed SMBBH sample with observation time as 30 years, in which four types of line stand for different confidence levels of 10/50/80/99 per cent. Orange circles stand for the distributions from the lensed SMBBH sample with observation time as 10 years. The magnitude of host galaxy is in F158 filter of RST which is estimated from the JAGUAR catalog \citep{Williams=2018}.
}
\label{fig:f8}
\end{figure}

Figure~\ref{fig:f8} shows the distributions of the galaxy mass and galaxy magnitude in the F158 filter of RST for the lensed SMBBHs detected by CPTA-opt (upper panel) and SKA-opt (lower panel) with $\rho \gtrsim 3$. The distributions obtained for the cases with a PTA observation period of $10$\,years (orange circles) and $30$\,years (green contour) are similar, in which galaxy masses $M_{\rm gal}$ are in the range of $\sim 10^{10} - 10^{12}M_{\odot}$ and galaxy magnitudes $m_{\rm gal}$ in the F158 filter of RST are in the range of $\sim 17 - 23$, substantially brighter than the limiting magnitude of RST/F158 $m_{\rm lim} \sim 26.8$. 

Note here that the sky coverage of RST is not considered in our estimation. The High Latitude Wide Area Survey of RST will cover $\sim 2,000\ \rm{deg}^2$ sky region. This means that the realistic detectable probability of lensed host galaxy by RST may decrease by a factor of $\sim 20$. However, there will be more than one sky survey carried out in the future, which may work complementarily to achieve much larger sky coverage. For instance, the Euclid \citep{Euclid2020} has the H-band limiting magnitude $m_{\rm lim,H}=24.5$, angular resolution $s\sim 0.3$\,arcsec and sky coverage $\sim 15,000\ \rm{deg}^2$; and China Space Station Telescope \citep[CSST;][]{Gong=2019} has the z-band limiting magnitude $m_{\rm lim,z}=24.1$, angular resolution $s\sim 0.18$\,arcsec and sky coverage $\sim 17,500\ \rm{deg}^2$. Furthermore, RST and other sky survey telescopes may be also used to specifically search for the lensed host of a PTA detected SMBBH in its localized sky area. With the operation of all these telescopes in the future, the all sky lensed host galaxies may have the chance to be observed. 

In addition, the real morphology identification of lensed extended galaxies is far more complicated than the simple criterion adopted in this paper and may require visual inspection. The detection strategy and science goal for each telescope may also affect the performance in this aspect. A more precise prediction for any one of those telescopes require more dedicated research.

To close the discussion of all the uncertainties in finding the lensed hosts of PTA SMBBHs above, we take the value of $14\%$ obtained above as the upper limit for the fraction of lensed hosts that can be found by galaxy sky surveys.

\subsection{Identification of lensed nHz GW signals}

We note here that associating the detected nHz GW signal with its lensed host galaxy is a difficult task even both of them are detected. In order to determine whether the nHz GW signal is lensed or not, one may have to match not only the localization area of the GW signal with the lensed host galaxy but also the properties of the SMBBH inferred from the GW signal and those inferred from the EM observations. For those non-active SMBBHs, only their lensed host galaxies but not themselves can be detected by deep and high-resolution galaxy surveys. Whereas, for those active SMBBHs, the multiple images of the active galactic nuclei (AGN) may be detected together with their lensed host galaxies. Below, we discuss the possible way to identify the lensed nHz GW signals from either non-active or active SMBBHs.

\subsubsection{Lensed non-active SMBBHs}

The SMBBHs generated from mergers of gas poor galaxies, which may be the dominant loud nHz GW sources, are normally inactive. The localization of the nHz GW signal from an SMBBH may reach $\sim 1$\,deg$^2$ \citep[e.g.,][]{Lee=2011, Guo=2020} and the number of lensed galaxies in such an area can be up to several tens \citep[e.g.,][]{YuH=2020}. Even the GW signal is known to be lensed, one has to figure out which lensed galaxy in the localization area of the GW signal is the corresponding host.
Furthermore, from the GW signal itself one cannot know whether it is lensed or not. In principle, one may first find the central non-active MBHs or SMBBHs of the lensed galaxies in the localization area via kinematics and dynamics by future sufficiently high resolution astrometric and spectral observations (e.g., by the Thirty Meter Telescope, Giant Magellan Telescope, and European Extremely Large Telescope; \citealt{Michalowski=2021}) and measure the properties of these MBHs or SMBBHs, though almost impossible by current telescopes. If the measurements can be done, then one could match the SMBBH properties with those determined from the GW signals and identify the lensed SMBBHs and its associated lensed host galaxies. 

\subsubsection{Lensed active SMBBHs}
\label{sec:5.2}

Gas-rich mergers of galaxies may lead to the formation of active SMBBHs appearing as AGNs, and it may contribute a significant fraction, $\sim 25\%$ \citep{Casey-Clyde=2022}, to the nHz GW sources.
Without morphology constraint, most of these lensed AGNs should be detectable in principle as they are one of the brightest sources in the universe and most of their multiple images are also resolvable\footnote{The population of those active SMBBHs has redshift peaked around $z\sim1$ \citep{Casey-Clyde=2022}. For similar galaxy-scale lensed point sources, BNSs which also have redshift peaked around $z\sim1$, \citet{MaHao=2023} predict that $\gtrsim 90$ per cent of multiple images are resolvable by RST.} by high-resolution space telescopes, such as HST and RST. We do not intend to go to details about the detectability of the active SMBBHs and its lensed host galaxies in this paper as our SMBBH sample is generated from gas-poor mergers. Nevertheless, we note that once the active SMBBHs are found by EM observations, one may estimate the properties of the SMBBHs by their various EM signatures \citep[][see also \citealt{Rodriguez=2006,Valtonen=2008,Dotti=2009,Morganti=2009,Eracleous=2012,Yan=2014,2015ApJ...809..117Y,Graham=2015,Charisi=2016,Liu=2016,Bansal=2017,D'Orazio=2018,Guo=2019,Laine=2020}, etc.]{Tamara=2022}. Then, by matching the position and physical properties of the lensed active SMBBHs with the nHz GW individual sources, one may find whether the nHz GW signals are associated with a lensed active SMBBH and thus identify the lensed nature of the GW signals.

\section{Conclusions}
\label{sec:6}

In this paper, we investigate the diffractive lensing effect on the nHz GWs emitted from SMBBHs lensed by intervening galaxies, and estimate the detectable number of such systems by future PTAs, including CPTA and SKA-PTA. We find that the diffractive lensing caused by the intervening galaxies lead to a small amplification of the GW amplitude ($\sim 1.04-1.14$ for $16-84\%$ of the mock sample and all $\lesssim 1.5$) and a small phase shift. The lensed GW signals are indistinguishable from those by similar SMBBH systems with slightly larger chirp masses or smaller distances without lensing. We estimate that future PTA experiments, such as CPTA and SKA-PTA, may detect about $10^2 - 10^4$ and $10^4 - 10^6$ individual SMBBHs within an observation period of $10-30$\,years, among which $\sim 0.01\%$ are strongly lensed by foreground galaxies, i.e., up to $\sim 106$ for CPTA and up to $\sim 289$ for SKA-PTA. The lensed nHz SMBBHs may be detectable by future PTAs but difficult to be identified as lensed ones through GW observations only. We further estimate that the lensed host galaxies of up to $14\%$ lensed nHz SMBBHs may be detected by sky survey telescopes, such as RST, and argue that the lensed active galactic nuclei hosting SMBBHs with nHz GW signals, contributing a significant fraction to the nHz SMBBHs ($\sim 25\%$; \citealt{Casey-Clyde=2022}), can be detected by electromagnetic telescopes. With the information from electromagnetic observations and measurements on the physical parameters of these lensed SMBBH systems, it may be possible to associate the nHz GW signals with a lensed non-active or active galaxy hosting an SMBBH and identify the diffractive lensing nature of the GW signals.   

\section*{Acknowledgements}
We thank Prof. Qingjuan Yu for helpful discussions and contribution to this paper.
This work is partly supported by the National Key Program for Science and Technology Research and Development (Grant nos. 2022YFC2205201, 2020YFC2201400), the National Natural Science Foundation of China (Grant nos. 12273050, 11690024, 11991052), and the Strategic Priority Program of the Chinese Academy of Sciences (Grant no. XDB 23040100).

\section*{Data Availability}

The data underlying this article will be shared on reasonable request to the corresponding author.



\bibliographystyle{mnras}
\bibliography{reference} 




\appendix



\bsp	
\label{lastpage}

\end{document}